\documentstyle[12pt,epsfig,amstex,amssymb]{article}

\begin{document}

\title{Unifying Nucleon and Quark Dynamics at Finite Baryon Number Density}

\author{%
J. Meyer$^{1}$,
K. Schwenzer$^{1}$
and
H.-J. Pirner$^{1,2}$ 
 \\[2mm]
{\small $^1$ Institut f\"ur Theoretische Physik der
Universit\"at Heidelberg,}\\ 
{\small D-69120 Heidelberg, Germany} \\
{\small $^2$ Max-Planck-Institut f\"ur Kernphysik, D-69029 Heidelberg, 
        Germany}}

\date{\today}

\maketitle

\centerline{PACS: 21.30.Fe , 21.65.+f , 21.60.-n}

\begin{abstract}
We present a model of baryonic matter which contains free constituent quarks in
addition to bound constituent quarks in nucleons. In addition to the common
linear $\sigma$-model we include the exchange of vector-mesons. The percentage
of free quarks increases with baryon density but the nucleons resist a
restoration of chiral symmetry.
\end{abstract}

\newpage

\section{Introduction}
The possibility of creating high density baryonic matter has triggered several
speculations about the appearance of a quark phase and/or diquark phase with
or without chiral symmetry restoration. Most naive quark models with effective
interactions have an inherent very low scale at which chiral symmetry is
restored. This scale can be obtained from an effective potential calculation
within a linear sigma model for example, and varies between 0.5 and 1.5 times normal nuclear density $\rho_0$.
This density appears to be unrealistically low, because higher mass mesons
(e.g. $\omega$-mesons) are neglected in these models. The inclusion of higher
mass mesons yields repulsive forces that drive the system away from the
chiral symmetry restoration phase transition.  \\
Relativistic Quantum Hadron Dynamics (QHD) has been a very successful mean
field theory of nuclear matter for certain features of nuclei. 
It gives a description of nuclear saturation and of the spin-orbit
potential in nuclei, which is close to that obtained phenomenologically. On the
other hand, its compressibility $K_\infty > 500$ MeV is too high, compared to
the empirical value $K_\infty = 210$ MeV \cite{blaizot}.
Although due to confinement nucleon degrees of freedom replace quark
degrees of freedom in vacuum, a growing part of the nucleon wave functions
will overlap with increasing nuclear density, because of the finite size of the
nucleons. By this, the substructure of the nucleon matters already before any
phase transition, and so it is interesting to ask the question what
chemistry of nuclear matter results if one combines both nucleons and
quarks. How is baryonic matter built up when a fraction of baryonic charge can
be delocalized in quarks, which have their own dynamics? In the seventies
various schemes of covalent bonding were invented. The above idea goes far
beyond these. It assumes the existence of an unconstrained quark substrate.

\section{Quark nucleon model}
We unify both quark and nucleon dynamics by a common linear $\sigma$-model
Lagrangian via which both sectors communicate. In equilibrium we do not need
complicated transition terms n $\leftrightarrow$ 3q. \\
The Lagrangian of our model reads
\begin{align}
  \label{eq:lagrangian}
  {\cal L} &= {\bar\psi}_n (i \gamma^\mu D_\mu^{(vn)} - g_{sn} (\sigma + i
  \vec\tau \vec\pi \gamma_5))\psi_n + {\bar\psi}_q (i \gamma^\mu \partial_\mu
  - g_{sq} (\sigma + i \vec\tau \vec\pi \gamma_5))\psi_q \nonumber \\
  & \qquad + \frac{1}{2} \left ((D_\mu^{(sv)} \vec\pi)^2 +
  (D_\mu^{(sv)} \sigma)^2 \right) 
  - \frac{1}{4} \Omega_{\mu \nu} \Omega^{\mu \nu}
  - {\cal V} (\sigma^2 + {\vec \pi}^2) \, , \nonumber
\end{align}
with the covariant derivatives $D_\mu^{(vn/sv)} = \partial_\mu + i g_{vn/sv}
\omega_\mu$, the field tensor $\Omega_{\mu \nu} = \partial_\mu \omega_\nu -
\partial_\nu \omega_\mu$ and the spontaneous symmetry breaking potential
($m_s^2 < 0$) 
\begin{equation}
  \label{eq:potential}
  {\cal V} (\sigma^2 + {\vec \pi}^2) = \frac{m_s^2}{2} (\sigma^2 + {\vec
  \pi}^2) + \frac{\lambda}{4} (\sigma^2 + {\vec \pi}^2)^2 \, . \nonumber
\end{equation}
The free constituent quarks with the same mass as quarks bound in nucleons
represent larger subclusters of quarks, for which a definite baryonic center
point can no longer be localized. These "free" quarks mimic the frequent
switching of string configurations. In this picture, the $\omega$-repulsion
is identified with the repulsion of the string junctions. \\
The hadronic part is a modification of the model presented by Walecka
\cite{walecka}, with the explicit mass term of the vector meson replaced
by a coupling to the scalar mesons. Thus in mean field theory, the $\omega$
also gains its mass from the $\sigma$-field. This leads to a scaling of
all hadron masses as suggested by Brown and Rho \cite{scaling}, who have also
investigated the compatibility of nucleon and quark degrees of freedom at
finite temperature and density \cite{buballa}. As a consequence, the chiral
symmetric phase with $\langle \sigma \rangle \approx 0$, i.e. very small
$\omega$-mass, is suppressed, because in the mean field approximation the
$\omega$-repulsion becomes very large. \\  
We choose the following couplings: $g_{sq} = 3.23$, $g_{sn} = 3\cdot g_{sq}
= 9.69 $, $m_s^2 = - 0.26$, $\lambda = 30.0$, $g_{sv} = 8.41$, $g_{vn} = 9.5$. 
The couplings of the mesonic potential $m_s^2$ and $\lambda$ are calculated
from a renormalization group (RG) 
flow-equation approach for the quark part of the model \cite{jochen}. 
The couplings yield a $\sigma$-mass of $m_\sigma$ = 720 MeV and a nucleon mass
of $m_n = 901$ MeV. The nucleon mass is less than 938 MeV because an explicit
symmetry breaking term is not included in our Lagrangian.
In the  mean field approximation with constant expectation values $\bar\sigma$
and $\bar\omega$, the Lagrangian takes the form
\begin{align}
  \label{meanfieldlagrangian}
  {\cal L}_{MF} &= \bar\psi_n (i \gamma^\mu \partial_\mu - g_{vn} \gamma^0
  \bar\omega - g_{sn} \bar\sigma) \psi_n
  + \bar\psi_q (i \gamma^\mu \partial_\mu - g_{sq} \bar\sigma) \psi_q
  \nonumber \\ 
  & \qquad +\frac{1}{2} g_{sv}^2 \bar\sigma^2 \bar\omega^2
  - \frac{m_s^2}{2} \bar\sigma^2
  - \frac{\lambda}{4} \bar\sigma^4 \, . \nonumber
\end{align}
The number of internal degrees of freedom is $\gamma_n$ = 4 for the nucleons
and $\gamma_q$ = 12 for the quarks, because of 2 spin and 2 isospin states
for both and in addition 3 different colors for the quarks.
In the ground state, the nucleons and quarks build two Fermi gases
with Fermi momenta $k_{Fn}$ and $k_{Fq}$. They are connected to the conserved
overall baryon density $\rho_B = \psi_n^\dagger \psi_n + \frac{1}{3}
\psi_q^\dagger \psi_q$ through 
\begin{equation}
  (1 - x) \rho_B = \frac{\gamma_n}{6 \pi^2} k_{Fn}^3 \nonumber \quad
  {\mathrm and} \quad
  3 x \rho_B = \frac{\gamma_q}{6 \pi^2} k_{Fq}^3 \, , \nonumber
\end{equation}
with $x$ denoting the concentration of quarks in the system. \\
Now one can calculate the energy density, which is\footnote{The
  first term in the second line is the same as the nucleon term in the first
  line with interchanged momenta and couplings}
\begin{align}
  \label{eq:energiedichte}
  \epsilon &= \frac{\gamma_n}{8 \pi^2} \left( \sqrt{k_{Fn}^2 + g_{sn}^2
  \bar\sigma^2} \left( k_{Fn}^3 + \frac{g_{sn}^2 \bar\sigma^2 k_{Fn}}{2}
  \right) - \frac{g_{sn}^4 \bar\sigma^4}{2} \log \left( \frac{k_{Fn} +
  \sqrt{k_{Fn}^2 + g_{sn}^2 \bar\sigma^2}}{g_{sn} \bar\sigma} \right) \right)
  \nonumber \\ 
  & \qquad + \frac{\gamma_q}{8 \pi^2} \left(k_{Fn} \leftrightarrow k_{Fq},
  g_{sn} \leftrightarrow g_{sq} \right)
  + \frac{\gamma_n^2 g_{vn}^2}{72 \pi^4 g_{sv}^2 \bar\sigma^2} k_{Fn}^6 +
  \frac{m_s^2}{2} \bar\sigma^2 
  + \frac{\lambda}{4} \bar\sigma^4 \, , \nonumber
\end{align}
where the mean field value $\bar\sigma$ in the ground state is
self-consistently given by minimizing $\epsilon$ with respect to $\bar\sigma$.

\section{Results}
\begin{figure}[ht]
  \centerline{\epsfig{file=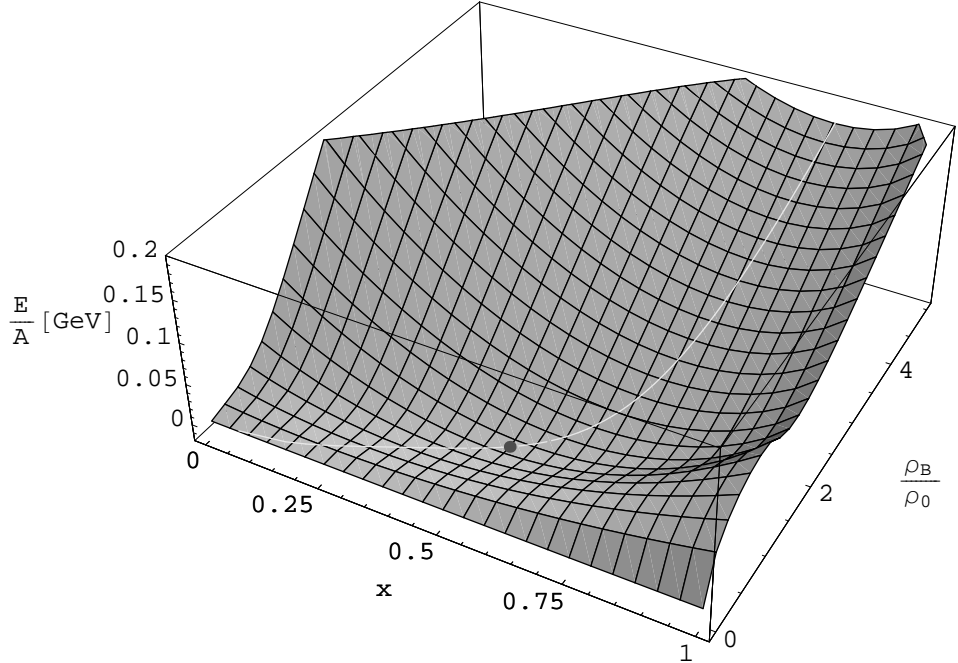,bbllx=100,bblly=440,bburx=445,bbury=720}}
%  \centerline{\epsfig{file=energysurface.eps}}
  \caption{Binding energy per baryon
    $\frac{E}{A} = \frac{\epsilon(\rho_B)-\epsilon(0)}{\rho_B} - m_N$ as a
    function of the independent variables $x$ and $\rho_B$ given in units of
    normal nuclear matter density $\rho_0 = 0.17$ fm$^{-3}$}
  \label{fig:energysurface}
\end{figure}
\begin{figure}[ht]
  \centerline{\epsfig{file=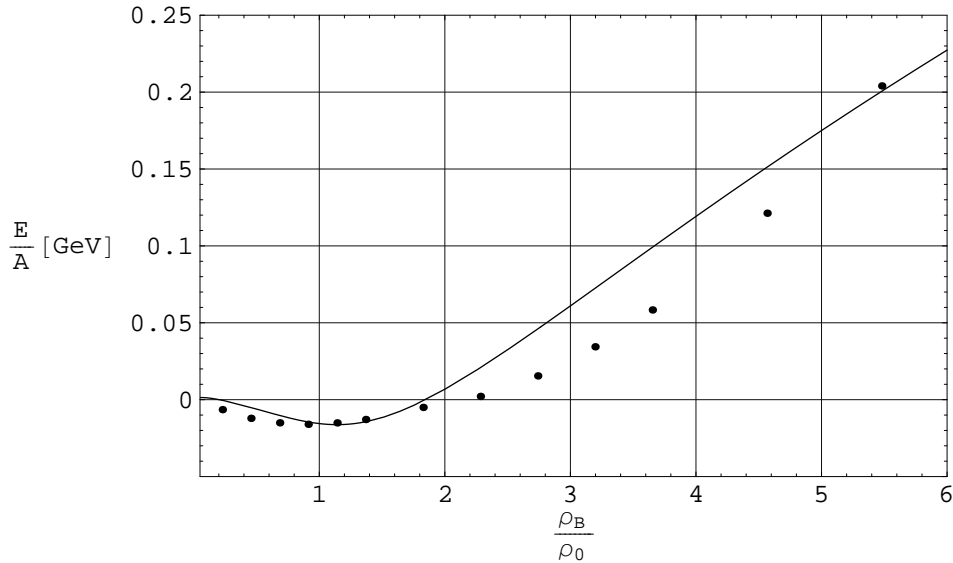,bbllx=100,bblly=495,bburx=445,bbury=720}}
%  \centerline{\epsfig{file=energycurve.eps}}
  \caption{Nuclear matter equation of state}
  \label{fig:energycurve}
\end{figure}

The result of our calculation is plotted in Fig.\ref{fig:energysurface}. To
the left the $x = 0$ boundary curve shows the pure hadronic phase. The
right boundary at $x = 1$ shows the same for pure quark matter with a kink due
to the first order chiral phase transition.
This transition persists only for a quark concentration $x
\gtrsim 0.99$, where at the endpoint the phase transition is of second 
order. The white line denotes the stable configuration with minimal energy at
each density and shows a smooth transition between a vanishing quark
concentration in vacuum and an asymptotic quark concentration $x_\infty
\approx 0.9$ as $\rho_B \rightarrow \infty$. \\
This curve of the energy minimum, which represents the equation of state (EOS)
of nuclear matter, is also plotted as a 2D projection in
Fig.\ref{fig:energycurve} together with the advanced nuclear physics
calculations by Pandharipande et. al. \cite{comparison}. Whereas quark matter
and nuclear  matter alone do not bind, the mixture of both binds at a
saturation density $\rho_S = 1.14 \rho_0$ with $E_B = 16$ MeV.

\begin{figure}[ht]
  \centerline{\epsfig{file=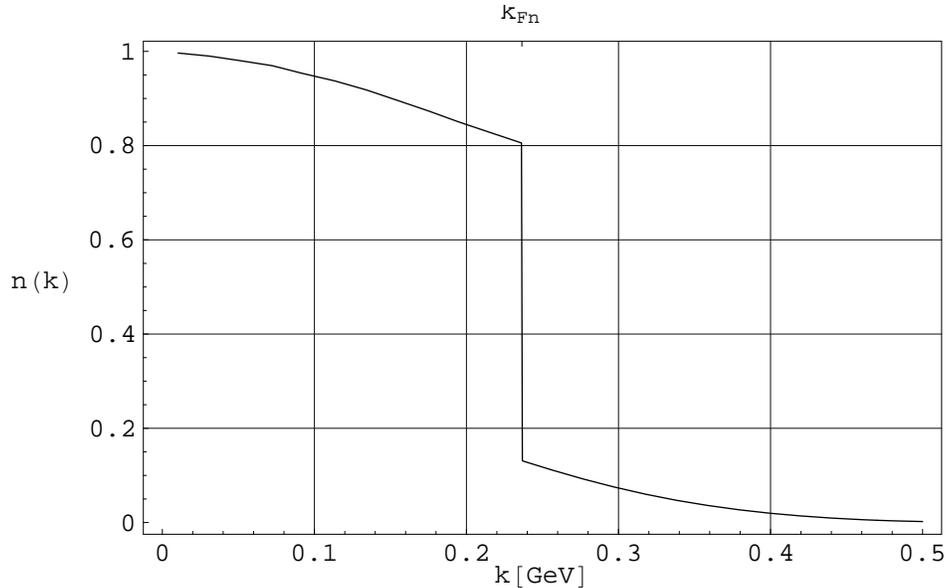,bbllx=100,bblly=495,bburx=445,bbury=720}}
%  \centerline{\epsfig{file=occupancy.eps}}
  \caption{Occupation numbers n(k) of quasi baryons in lead}
  \label{fig:occupancy}
\end{figure}

The Dirac mass of the nucleon at saturation is $m_D^* = g_{sn}
\bar\sigma_s = 581$ MeV. As pointed out in \cite{effmass} this quantity is not
the relativistic analogon of the nonrelativistic effective mass measured in
nuclear experiments. The corresponding quantity is the relativistically
equivalent effective mass $m_e^*$. We get $\frac{m_e^*}{m_n} = 1 -
\frac{g_{vn} \bar\omega}{m_n} = 0.71$ compared to an experimental value of
$\frac{m^*}{m_n} \approx 0.74$ \cite{effmass}. \\
As in the standard Walecka model the compression modulus $K_\infty =823$ MeV
is much too large. This is a general problem of these simple mean field models
and could be corrected by adding further meson interactions or potential
terms to the Lagrangian. \\
The saturation value for the quark concentration is $x_s = 50 \% $. 
This rather high value is reduced in finite nuclei due to the large
surface of low nuclear matter density. A simple estimation with a Fermi
distribution for the density gives $\langle x \rangle \approx 40 \%$ for
$^{208}$Pb and $\langle x \rangle < 30 \% $ for $^{16}$O.
There are indeed experiments on the occupancies of lower shell model orbitals,
which suggest that these orbitals are not entirely filled. Instead the Fermi
surface is smeared out. This effect has been explained by a mixing of shell
model orbitals, but in our model it is simply due to the coupling of different
quark momenta to a baryonic three quark system. Such a composite baryon is
not necessarily a nucleon but may be another excited baryonic state. 
Fig.\ref{fig:occupancy} shows the occupancies in lead computed with this
model and looks almost the same as the result obtained under the inclusion of
ground state correlations \cite{hasse}. In the model described here, the
"free" constituent quarks reduce the amount of quasi baryons below the Fermi
momentum and enhance their number with momenta higher than $k_{Fn}$.
The unchanged part of the Fermi surface is connected to the residue factor $z$,
which can be measured. Because the 3s-orbital in lead is nearly exclusively
located in the center of the nucleus, one can take in very good
approximation the nuclear matter value of the quark concentration to compute
the residue factor for this orbital. The result of $z = 0.67$ compares
favorably to the value $z = 0.64 \pm 0.06$ from (e, e$^\prime$p) experiments \cite{sick}.

\section{Conclusion}
Because of the repulsive $\omega$-mean field at saturation $g_{vn}
\bar\omega_s = 263$ MeV, the nucleons occupy the higher energy levels,
whereas the constituent quarks have a strong occupancy in the deep lying
levels. In  such a way the overlap of constituent quarks and nucleons is
restricted to the upper levels and would lead to a small correction of the
energy density. Note that this overlap remains small at densities regarded
here. The validity of our calculation is also limited by the credibility of
the mean field 
approximation and the appearance of gluonic degrees of freedom at high
density. We estimate the momentum scale for this region to be $k_{UV} \geq 1$
GeV cf. with our RG calculation \cite{bj}.   
Due to the nonabelian nature of QCD the three body
gluon exchange forces are more important than the two body gluon
exchanges. The three body forces lead to baryonic tripling whereas two body
forces induce diquark formation \cite{diquark}. The latter may be an
intermediate stage on the way from high density to low density, but as shown
in nature color neutralization will win ultimately at lower densities. \\ \\
To summarize, we have constructed a very simplified model with constituent
quarks and nucleons, which interpolates between low density nucleon matter and
high density quark matter in a continuous way. As a result we have shown that
our model qualitatively reproduces the nuclear matter EOS and provides a simple
explanation for experimental measured residue factors.


\begin{thebibliography}{99}
\bibitem{blaizot}
  J.P. Blaizot, Phys. Rep. {\bf 64} (1980) 171
\bibitem{walecka}
  J.D. Walecka, Theoretical Nuclear and Subnuclear Physics, \\
  Oxford University Press (1995)
\bibitem{scaling}
  G.E. Brown and M. Rho, Phys. Rev. Lett. {\bf 66} (1991) 2720
\bibitem{buballa}
  G.E. Brown, M. Buballa and M. Rho, Nucl. Phys. A {\bf 609} (1996) 519
\bibitem{jochen}
  T. Kunihiro, J. Meyer, G. Papp and H.J. Pirner, to be published
\bibitem{comparison}
  A. Akmal, V.R. Pandharipande, D.G. Ravenhall, \\
  Phys. Rev. C {\bf 58} (1998) 1804 
\bibitem{effmass}
  M. Jamion, C. Mahaux, Phys. Rev. C {\bf 40} (1989) 354
\bibitem{hasse}
  R.W. Hasse, B.L. Friman and D. Berdichevsky, \\
  Phys. Lett. B {\bf 181} (1986) 5  
\bibitem{sick}
  I. Sick and P.K.A. de Witt Huberts, \\
  Comm. Nucl. Part. Phys. {\bf 40} (1991) 177
\bibitem{bj}
  B.J. Sch\"afer and H.J. Pirner, hep-ph/9903003, subm. to Nucl. Phys. A
\bibitem{diquark}
  M. Alford, K. Rajagopal and F. Wilczek, Phys. Lett. B {\bf 422} (1998) 247
\end{thebibliography}
\end{document}